# A method for accelerating projection-based magnetic particle imaging

Kenya Murase

*Department of Medical Physics and Engineering, Faculty of Health Science, Graduate School of Medicine, Osaka University, Suita, Osaka, Japan*

*Department of Future Diagnostic Radiology, Graduate School of Medicine, Osaka University, Suita, Osaka, Japan*

Magnetic particle imaging (MPI) is an imaging method that can visualize magnetic nanoparticles in positive contrast, without radiation exposure. Recently, we proposed an image reconstruction method for projection-based MPI (pMPI), in which the system function was incorporated into the simultaneous algebraic reconstruction technique and the total variation minimization was used to suppress noise and artifacts. This study investigated the usefulness of our method for accelerating pMPI through simulation and phantom experiments with varying number of projections. The present results suggest that our method is useful for accelerating pMPI without deteriorating the image quality.

Magnetic particle imaging (MPI) is an imaging method that utilizes the non-linear response of magnetic nanoparticles (MNPs) to an external alternating magnetic field,[1] and can visualize MNPs in positive contrast, without radiation exposure.[2,3] Recently, we proposed a method for simultaneous correction of sensitivity and spatial resolution in projection-based MPI (pMPI).[4] In this method, the system function (SF) was incorporated into the simultaneous algebraic reconstruction technique (SART)[5] and total variation (TV) minimization was used as a regularizer to suppress noise amplification and artifacts. For the practical application of pMPI, it is necessary to shorten the data acquisition time as much as possible. There are two methods for accelerating pMPI. One is to shorten the data acquisition time for each projection, and the other is to reduce the number of projections ($N_p$) or combination of both. This study aimed to investigate the effect of $N_p$ on the quality of pMPI images obtained by our method and to evaluate the usefulness of our method for accelerating pMPI through simulation and phantom experiments.

The details of our method are described in our previous paper.[4] Briefly, the image reconstruction was performed iteratively based on the following two procedures: First, the image at pixel $j$ and iteration $n$ ($f_j^n$) was updated as follows:

$$f_j^{n+1} = f_j^n + \frac{\lambda_n}{\sum_i a_{ij}} \sum_i \left( \frac{g_i - \sum_k a_{ik} f_k^n}{\sum_k a_{ik}} \right) a_{ij} \quad (n = 1, 2, \cdots, N \text{ and } j = 1, 2, \cdots, J), \quad (1)$$

where $\lambda_n$ denotes the relaxation parameter at iteration $n$, which was fixed at 1.0, $g_i$ the projection data at bin $i$, $N$ the number of iterations, and $J$ the total number of pixels. The term $a_{ij}$ denotes the element of the system matrix, which is expressed as[4]

$$a_{ij} = S(j) \cdot SF(d_{ij}), \quad (2)$$

where $S(j)$ is the sensitivity of the receiving coil at pixel $j$, and $SF(d_{ij})$ denotes the SF value at distance $d_{ij}$ between pixel $j$ and the field-free line (FFL) at bin $i$.[4] In this study, the SF was measured using a point source as described afterwards, and $S(j)$ was calculated using the Biot–Savart law based on the principle of reciprocity.[6]

After the SART update using Eq. (1), a regularization method with TV minimization was used. For the TV minimization, isotropic TV, that is, the sum of the gradients of the image[7] was used, and the gradient descent method was applied on $f_j^{n+1}$, with $f_j^{n+1,1}$ set as $f_j^{n+1}$ obtained in the first step,[4] as shown in Eq. (3).

$$f_j^{n+1,m+1} = f_j^{n+1,m} + \alpha \cdot d_f \cdot \text{div}\left(\frac{\nabla f_j^{n+1,m}}{\left\|\nabla f_j^{n+1,m}\right\|_2}\right) \ (m = 1, 2, \cdots, M \text{ and } j = 1, 2, \cdots, J), \tag{3}$$

where $\alpha$ and $M$ denote a regularization parameter and the number of iterations for TV minimization, respectively, $d_f$ the amount of change in the first reconstruction step, div and $\nabla$ the divergence and gradient operators, respectively, and $\|\cdot\|_2$ the $\ell_2$ norm. After $m$ reached $M$, Eq. (1) was applied again, with $n$ and $f_j^n$ set as $n+1$ and $f_j^{n+1,M+1}$, respectively.

The iterative procedure above was repeated until $n$ reached $N$ or $\sqrt{\sum_{j=1}^{J}(f_j^{n+1} - f_j^n)^2} / \sqrt{\sum_{j=1}^{J}(f_j^n)^2} < \varepsilon_{tol} \ (n \geq 2)$ was satisfied. In this study, $N$, $M$, $\alpha$, and $\varepsilon_{tol}$ were set to 1000, 1, 0.05, and $10^{-4}$, respectively. The initial estimate $f_j^1 \ (j = 1, 2, \cdots, J)$ was set to a uniform image with zero pixel intensity. When the reconstructed image contained negative values, these values were set to zero.

In the simulation, a vortex-shaped numerical phantom with an image matrix size of 128×128 was used. First, projection data with various $N_p$ were generated using the forward projection method and were convolved with the SF. In addition, Gaussian noise with a noise level of 5% was added to the projection data using normally distributed random numbers. Here, the noise level is defined as the standard deviation of the noise divided by the maximum value of the projection data. Image reconstruction from the generated projection data was performed using our method. For comparison, the filtered back-projection (FBP) method with a Shepp–Logan filter[8] was also used. As in our method, when the image reconstructed using the FBP method contained negative values, these values were set to zero. The $N_p$ over 180° was varied between 4 and 180. The angular increment between two successive projections is $180/N_p$ (°).

The reconstructed images were quantitatively evaluated using two measures. The first measure is the percent root mean square error (PRMSE), which was calculated as follows:

$$PRMSE = \sqrt{\frac{\sum_{j=1}^{J}(f_j^r - f_j^g)^2}{\sum_{j=1}^{J}(f_j^g)^2}} \times 100 \ (\%), \tag{4}$$

where $f_j^r$ and $f_j^g$ denote the image intensities at pixel $j$ of the reconstructed and ground truth (GT) images, respectively. The GT image is shown in the upper row of Fig. 2(a).

The other measure is the structural similarity (SSIM) index, proposed by Wang et al.[9]. It was calculated as follows:

$$SSIM = \frac{(2\mu_r\mu_g+c_1)(2\sigma_{rg}+c_2)}{(\mu_r^2+\mu_g^2+c_1)(\sigma_r^2+\sigma_g^2+c_2)}, \tag{5}$$

where $\mu_r$ and $\mu_g$ denote the averages within the windows set on the reconstructed and GT images, respectively; $\sigma_r^2$ and $\sigma_g^2$ denote the corresponding variances; $\sigma_{rg}$ denotes the covariance between the two windows. The terms $c_1$ and $c_2$ denote two variables to stabilize the division with a small denominator. We used the default values given by Wang et al.[9] for the window size, $c_1$, and $c_2$. The resultant SSIM is a decimal value between −1 and 1, and a higher SSIM indicates higher image similarity. An SSIM of 1 implies that the two images are identical.

Phantom experiments were performed using an OU-shaped phantom and our MPI scanner.[10,11] The details of our MPI system have been reported in our previous papers.[10,11] Briefly, a selection magnetic field (SMF) was generated by two opposing neodymium magnets, and an FFL was generated at the center of the two neodymium magnets. The gradient strengths of the SMF perpendicular and parallel to the FFL were 3.9 T/m and 0.1 T/m, respectively. A drive magnetic field (DMF) for exciting the magnetization of MNPs was generated using a solenoid excitation coil (length: 100 mm, inner diameter (ID): 80 mm, and outer diameter (OD): 110 mm). The frequency and peak-to-peak strength of the DMF were 400 Hz and 20 mT, respectively. A gradiometer coil (length: 50 mm, ID: 35 mm, and OD: 40 mm) was placed in the excitation coil to receive the signal generated by MNPs. The third-harmonic signal was extracted using a lock-in amplifier and was converted to digital data by a multifunction data acquisition device.

To acquire projection data, the OU-shaped phantom (1.5-mm-ID silicon tubes filled with MNPs) placed in the receiving coil was automatically rotated around the axis of the receiving coil through 180° in steps of 1° ($N_p$ = 180) and translated in the direction perpendicular to the axis of the receiving coil from −16 mm to +16 mm in steps of 1 mm using an XYZ-axes rotary stage. The data acquisition time was 1 s per position. The projection data with different $N_p$ were generated by thinning out the projection data acquired above. Each projection data was then transformed into 64 bins by linear interpolation. In this study, Resovist® (Fuji Film RI Pharma Co., iron amount: 27.9 mg/mL) was used as the MNPs.

MPI images (matrix size: 64×64) were reconstructed from the projection data generated above using our method and the FBP method. To quantitatively evaluate the reconstructed images, the profiles along the horizontal line passing through the center of the reconstructed images were calculated.

The SF was measured using a point source (length: 1 mm and ID: 1 mm) filled with MNPs. The point source was placed at the center in the receiving coil and was translated in the direction perpendicular to the axis of the receiving coil from −10 mm to +10 mm in steps of 0.5 mm using the XYZ-axes rotary stage. The data acquisition time was 30 s per position.

Figure 1 shows the measured SF (red closed circles) and that fitted to the sum of two Gaussian functions (blue solid line). In this study, the fitted curve was normalized to the unit sum and was used as the SF in Eq. (2).

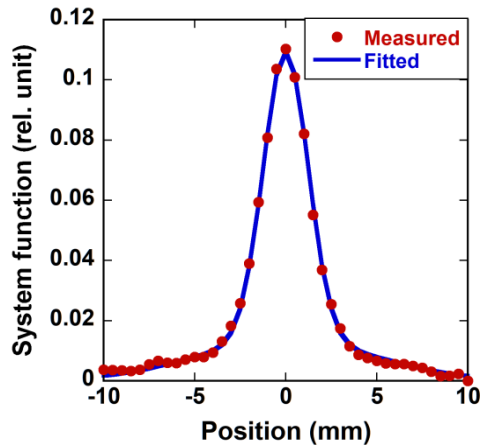

**Fig. 1** System function measured using a point source (red closed circles) and that fitted to the sum of two Gaussian functions (blue sold line).

Figure 2(a) shows the reconstructed images of the vortex-shaped phantom obtained by the FBP method (middle row) and our method (lower row) for different $N_p$ indicated at the bottom. As shown in Fig. 2(a), our method yielded the images with much less noise and artifacts compared to those obtained by the FBP method and similar to the GT image at $N_p \geq 12$.

Figure 2(b) shows the PRMSE values for the FBP method (red solid line) and our method (blue dotted line) as a function of $N_p$, whereas Fig. 2(c) shows the SSIM values for them. As shown in Fig. 2(b), the PRMSE for our method was significantly lower than that for the FBP method. When using the FBP method, the PRMSE gradually decreased with increasing $N_p$. In contrast, when using our method, the PRMSE decreased with increasing $N_p$ at $N_p < 12$, following which it became almost constant.

As shown in Fig. 2(c), our method produced much higher SSIM than the FBP method. When using the FBP method, the SSIM gradually increased with increasing $N_p$. In contrast, when using our method, it rapidly increased and had a peak at $N_p = 12$, following which it plateaued.

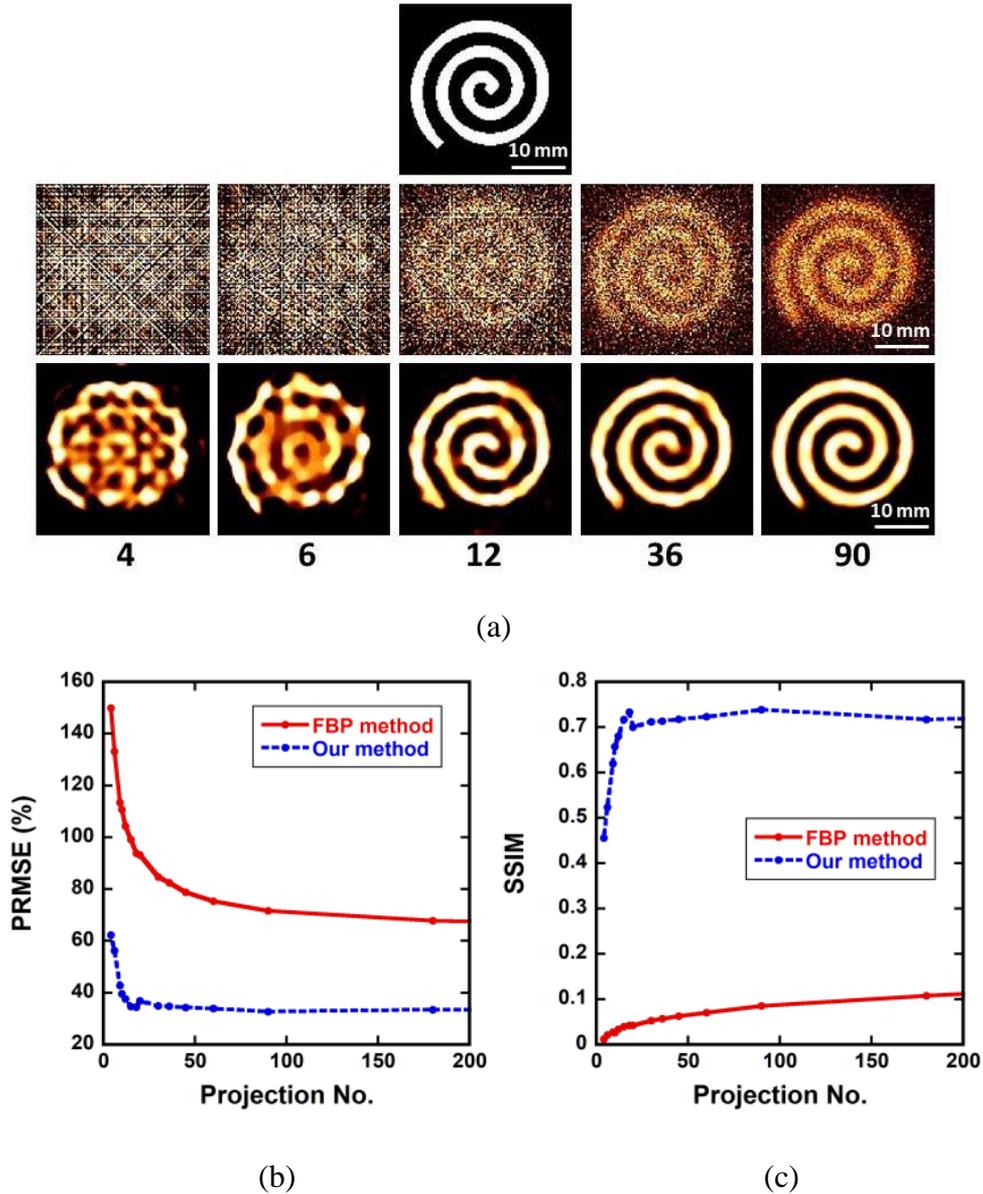

(a)

(b)                                          (c)

**Fig. 2** (a) Ground truth image of a vortex-shaped numerical phantom (upper row) and images reconstructed using the FBP method (middle row) and our methods (lower row) for different projection numbers ($N_p$) indicated at the bottom. The display window is set the same in all images. Scale bar = 10 mm. (b) PRMSE and (c) SSIM as functions of $N_p$ for the images reconstructed using the FBP method (red solid line) and our method (blue dotted line).

Figure 3(a) shows the images of the OU-shaped phantom reconstructed using the FBP method (upper row) and our method (lower row) for different $N_p$ indicated at the top. Figure 3(b) shows the profiles along the horizontal line passing through the center of the reconstructed images at $N_p = 6$ for the FBP method (red solid line) and our method (blue dotted line). As shown in Fig. 3(a), the quality of the images obtained by our method was significantly better than that for the FBP method. When using our method, artifacts and blurring almost disappeared even at $N_p = 6$. This was also confirmed by the profiles shown in Fig. 3(b).

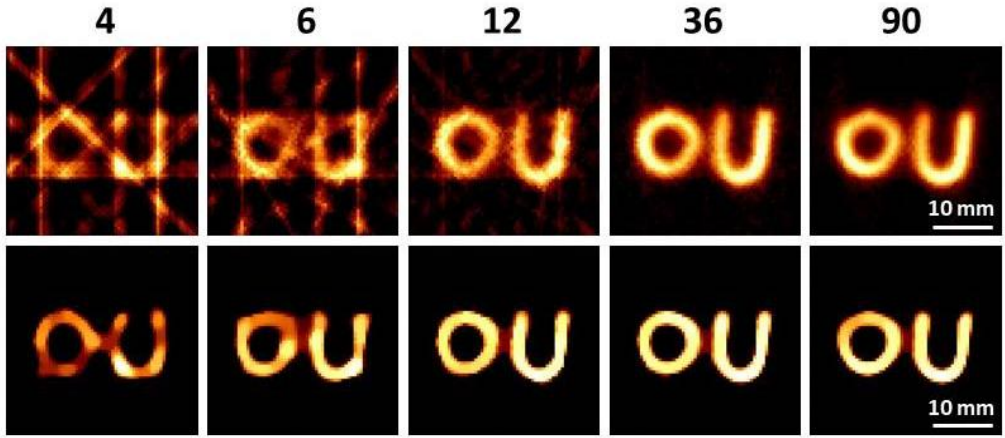

(a)

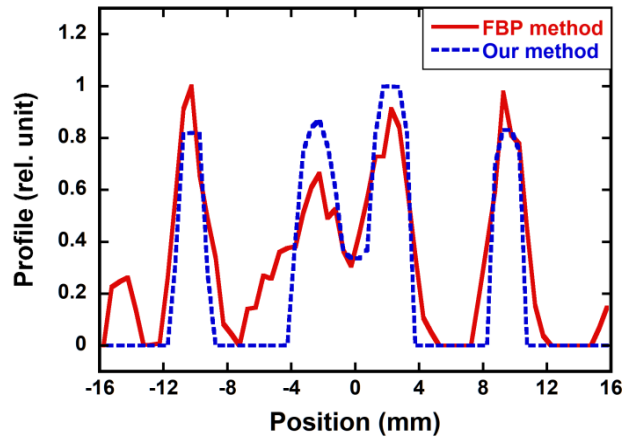

(b)

**Fig. 3** (a) Reconstructed images of an OU-shaped phantom obtained by the FBP method (upper row) and our method (lower row) for different $N_p$ indicated at the top. Scale bar = 10 mm. (b) Profiles along the horizontal line passing through the center of the images reconstructed using the FBP method (red solid line) and our method (blue dotted line) for $N_p$ = 6. The maximum intensities of the reconstructed images are normalized to unity in both methods.

In the simulation (Fig. 2), our method exhibited significant improvement in image quality and produced significantly lower PRMSE and higher SSIM than the FBP method even at extremely low $N_p$. The phantom experiments (Fig. 3) also demonstrated that the deterioration of image quality was significantly suppressed by our method even at extremely low $N_p$. Thus, these results suggest that our method is useful for accelerating pMPI.

Theoretically, the $N_p$ required for obtaining good reconstructed images can be estimated from the Nyquist–Shannon (sampling) theorem.[12] According to this theorem, a unique reconstruction of an object sampled in space is obtained if the object was sampled with a frequency greater than twice the highest frequency of the object details. Otherwise, it causes aliasing artifacts. In the parallel scan mode for computed tomography, the number of scanned points per one projection line ($N_s$) is required to satisfy the following relationship:[13]

$$N_p \geq \frac{\pi}{2} \times N_s. \tag{6}$$

$N_s$ is usually equal to the square root of image matrix size. When $N_s$ is 64 or 128, $N_p$ should be equal to or greater than 100 or 201, respectively, to satisfy the sampling theorem.[12] However, when using our method (Figs. 2 and 3), satisfactory images were obtained even if $N_p$ was much lower than that satisfying Eq. (6), whereas the images obtained by the FBP method were remarkably deteriorated and artifacts were observed (Figs. 2 and 3).

The TV minimization used in our method (Eq. (3)) is effective for inducing sparsity in image processing.[7] According to the theory of compressed sensing,[14] the sparsity of an image can be exploited to reconstruct it from far fewer projections than required by the sampling theorem.[12]. Equation (3) is equivalent to solving the diffusion equation used in the anisotropic diffusion method for image denoising,[15,16] in which the diffusion coefficient is proportional to the reciprocal of $\left\|\nabla f_j^n\right\|_2$. Thus, when $\left\|\nabla f_j^n\right\|_2$ decreases, the diffusion progresses, resulting in an increase in the piecewise smoothness of images. Conversely, when it increases, the diffusion decreases, resulting in the edge preservation of images. Thus, these features appear to be the main reason why the superiority of our method is maintained even at much lower $N_p$ than that satisfying Eq. (6).

In summary, the present simulation and phantom experiments demonstrated that our method is useful for improving the image quality of pMPI even at extremely low $N_p$. Thus, our method will be useful for accelerating pMPI.

**Acknowledgments**

The author thanks Mr. Shinichiro Morishita for his help in phantom experiments. This work was supported by Grants-in-Aid for Scientific Research (Grant Nos.: 25282131 and 15K12508) from the Japan Society for the Promotion of Science (JSPS).

**References**

1) B. Gleich and J. Weizenecker, Nature **435**, 1214 (2005).
2) B. Zheng, M. P. von See, E. Yu, B. Gunel, K. Lu, T. Vazin, D. V. Schaffer, P. W. Goodwill, and S. M. Conolly, Theranostics **6**, 291 (2016).
3) P. Chandrasekharan, Z. W. Tay, X. Y. Zhou, et al. Br. J. Radiol. **91**, 20180326 (2018).
4) K. Murase, Med. Phys. **47**, 1845 (2020).
5) A. H. Andersen and A. C. Kak, Ultrason. Imaging **6**, 81 (1984).
6) J. Rahmer, J. Weizenecker, B. Gleich, and J. Borgert. BMC Med. Imaging **4**, 1 (2009).
7) C. R. Vogel and M. E. Oman, IEEE Trans. Image Process. 7, 813 (1998).
8) L. A. Shepp and B. F. Logan, IEEE Trans. Nucl. Sci. **NS-21**, 21 (1974),
9) Z. Wang, A. C. Bovik, H. R. Sheikh, and E. P. Simoncelli EP. IEEE Trans. Image Process. **13**, 600 (2004).
10) K. Murase, S. Hiratsuka, R. Song, and Y. Takeuchi, Jpn. J. Appl. Phys. **53**, 067001 (2014).
11) K. Murase, R. Song, and S. Hiratsuka, Appl. Phys. Lett. **104**, 252409 (2014).
12) H. Nyquist, Trans. AIEE **47**, 617 (1928).
13) F. Kharfi, in *Imaging and Radioanalytical Techniques in Interdisciplinary Research - Fundamentals and Cutting Edge Applications*, ed. F. Kharfi (InTech, London, 2013) Chap. 4.
14) D. L. Donoho, IEEE Trans. Inf. Theory **52**, 1289 (2006).
15) P. Perona and J. Malik, IEEE Trans. Pattern Anal. Mach. Intell. **12**, 629 (1990).
16) K. Murase, Y. Yamazaki, M. Shinohara, K. Kawakami, K. Kikuchi, H. Miki, T. Mochizuki, and J. Ikezoe, Phys. Med. Biol. **46**, 2713 (2001).